# AN INITIAL PEER CONFIGURATION ALGORITHM FOR MULTI-STREAMING PEER-TO-PEER NETWORKS

Tomoyuki Ishii and Atsushi Inoie[*]

Department of Network Engineering, Kanagawa Institute of Technology,
Atsugi-city, Japan
`inoie@nw.kanagawa-it.ac.jp`

## ABSTRACT

*The growth of the Internet technology enables us to use network applications for streaming audio and video. Especially, real-time streaming services using peer-to-peer (P2P) technology are currently emerging. An important issue on P2P streaming is how to construct a logical network (overlay network) on a physical network (IP network). In this paper, we propose an initial peer configuration algorithm for a multi-streaming peer-to-peer network. The proposed algorithm is based on a mesh-pull approach where any node has multiple parent and child nodes as neighboring nodes, and content transmitted between these neighboring nodes depends on their parent-child relationships. Our simulation experiments show that the proposed algorithm improves the number of joining node and traffic load.*

## 1. INTRODUCTION

Real-time large-volume multimedia streaming services such as on-demand broadcasting and videoconferencing using peer-to-peer (P2P) technology have recently become commonplace. The major advantage of using P2P technology is reducing the load and the anticipated number of host servers because some users who receive video content share the server's role to transmit the content to other users.

In the P2P network, the construction of a logical network (overlay network) in a physical network (IP network) is a major issue. The roles of a video streaming server and a peer are like a parent and a child, respectively. Methods to construct a logical network are divided into the tree-push approach and the mesh-pull approach [1]. In the tree-push approach, a tree structure is constructed by assigning streaming servers as root nodes to transmit content from the root nodes to leaf nodes (peers). Therefore, its topological structure can be simple and its advantages include the possible suppression of an unstable delay in the transmission. OverCast [2] and ESM [3] are applications of this type of transmission method. In the mesh-pull approach, any node has multiple parent and child nodes as neighboring nodes, and content transmitted between these neighboring nodes depends on their parent-child relationships. Representative systems that use the mesh-pull approach include CoolStreaming [4] and Chainsaw [5].

One problem in the tree-push approach is that all the child nodes will lose reception of the video content if the upper-level peers depart from the system after completing the service. However, this issue can be avoided in the mesh-pull approach because various network topology configurations are constructed by switching to other parent nodes even if some nodes depart from the system. Therefore, the mesh-pull approach has attracted much attention. A drawback of a system based on the mesh-pull approach is that the implementation of simple algorithms, which are easily offered in the tree-push approach, is very difficult due to its complicated network topology. For example, CoolStreaming uses *SCAMP* [6], an algorithm that determines a parent-child relationship between peers when constructing network topology. SCAMP attempts to improve scalability for a growing number of viewers by stochastically selecting transmission destinations.

---

[*] Corresponding author.

Additionally, compared to a randomly constructed network, a logical network can be more streamlined by incorporating algorithms used in the tree-push approach into a system based on the mesh-pull approach. Fukushima et al. [7, 8] have proposed topology construction algorithms called *peer selection algorithms,* in a mesh-pull P2P streaming network. Based on the algorithms by [7, 8], Ishii and Inoie [9] also considered topology construction algorithms for a peer-to-peer streaming network where peer leaving occurs.

In this paper, we extend the algorithms proposed by [9] for applying a multi-streaming environment [10]. In a multi-streaming environment, some peers require multiple different video contents. Hence, we must consider multiple logical networks on a physical network simultaneously. Through some simulation experiments, we show that our extended algorithms are valid in the above-mentioned environment.

The remaining of this paper is as follows. In Section 2, we survey the related work on P2P streaming technologies. In Section 3, we describe our P2P streaming model. In Section 4, we propose an extended initial peer allocation algorithm. In Section 5, we show some simulation experiments. Finally, in Section 6, we conclude this paper.

## 2. RELATED WORK

P2P streaming has attracted much attention in recent years. In this section, we review some articles of P2P streaming technologies.

Liu et al. [11] considered an efficient P2P multi-streaming mechanism using a tree-push approach. Wu et al. [12] studied an analytical model of multichannel P2P live video systems. For evaluating the scalability of the systems, they used simple queueing network models. They numerically compared the performance between the single and multichannel P2P networks.

Park et al. [13] proposed an adaptive topology construction algorithm called Climber, which is based on the hybrid approach of a tree-push and a mesh-pull. In their study, it is assumed that the bandwidth for receiving content is always enough. They showed the effectiveness of their proposed algorithm via some simulation experiments.

Magharei et al. [14] proposed a tax-based contribution-aware scheme for mesh-pull P2P streaming approaches. In their scheme, a tax function is used to determine the number of parent peers. Xie et al. [4] proposed a mesh-pull P2P streaming system called CoolStreaming, which consists of three key modules: membership manager, partnership manager and scheduler. In CoolStreaming, SCAMP [6] is used as an initial topology construction algorithm. SCAMP choose parent and child peers randomly. Hence, it is difficult to guarantee the optimality of the algorithm.

The above mentioned papers, however, the initial topology construction methods of the logical networks did not discussed well.

Fukushima et al. [7, 8] have proposed topology construction algorithms in a P2P streaming network where all the origin streaming servers have the same video content. Their algorithms are called peer selection algorithms and are based on a mesh-pull approach. Their proposed algorithms are 1) a minimum logical hop (MLH) algorithm to increase the number of peers that can concurrently receive service and 2) a minimum physical hop (MPH) algorithm to reduce the physical traffic volume. They have demonstrated the characteristics of these algorithms using computer simulations.

In large-scale systems, MLH and MPH algorithms should be more efficient than SCAMP because the former two algorithms construct a logical network based on the structure of the physical network. However, Fukushima et al. [7, 8] did not consider the departure of peers that have finished viewing video, which is often encountered in real P2P networks. The departure of peers may extensively alter logical network topology, and lead to problems where nodes with narrow bandwidths are concentrated near the root node and physically remote nodes are connected with each other.

Ishii and Inoie [9] considered a P2P streaming network where peer leaving occurs. Based on the algorithms by [7, 8], they proposed two peer exchange algorithms: one to further increase the number of peers that can be concurrently connected by considering transmission bandwidth and the other to reduce physical traffic volume. They conducted simulation experiments to demonstrate that the proposed algorithms not only increase the number of concurrently connected peers and reduce physical traffic volume, but also effectively manage the departure of peers.

In this paper, we extend the algorithms proposed by [9] for applying a multi-streaming environment [10]. Indeed, we expect that our proposed algorithms lead an efficient solution to controlling P2P streaming networks. The algorithms can be also combined with existing P2P streaming such as CoolStreaming and Climber by replacing their initial topology construction algorithms.

## 3. P2P Streaming Network Model

Figure 1 shows the video streaming network model discussed herein. In this streaming network, $S$ nodes are assumed to be interconnected where $\mathcal{S} = \{1,2, \ldots, S\}$ represents a set of these nodes. This network is divided into multiple autonomous systems (ASs) and each node belongs to one AS. Herein ASs represent individual networks that Internet service providers (ISPs) or companies maintain and operate. The transmission source node of broadcasting is called the origin streaming server (OSS) and its set is defined as $\mathcal{S}_0$. All other nodes are called peers, and a set of these nodes is defined by $\mathcal{S}_1 = \mathcal{S} \setminus \mathcal{S}_0$. The total number of peers contained in this network is denoted by $S_1 = |\mathcal{S}_1|$.

The set of video contents is denoted by $\mathcal{K} = \{1,2, \ldots, K\}$. We assume that each OSS can provide peers with at most one kind of video content and assume that the network has $K$ OSSs (i.e., $|\mathcal{S}_0| = K$).

At a given time, each peer is either in service-receiving mode or waiting mode. Peers that are in service-receiving mode inquire a tracker server of a video streaming source peer (or OSS). The tracker server determines the initial peer that transmits video content using an initial peer configuration algorithm, which is discussed in the next section. The set of video contents required by peer $i$ is denoted by $\mathcal{K}(i)$, that is, some peers may require multiple different video contents. We denote by $\Pi(\mathcal{K})$ the power set (i.e., all subsets of $\mathcal{K}$), and define the probability that a peer requires the set $V(\subset \mathcal{K})$ of video contents such that $\sum_{V \in \Pi(\mathcal{K})} P(V) = 1$. A set of nodes that transmit video content $k$ to peer $i$ is defined by $\mathcal{P}^k(i)$, while a set of nodes that receive video content $k$ from a peer $i$ is defined by $\mathcal{C}^k(i)$.

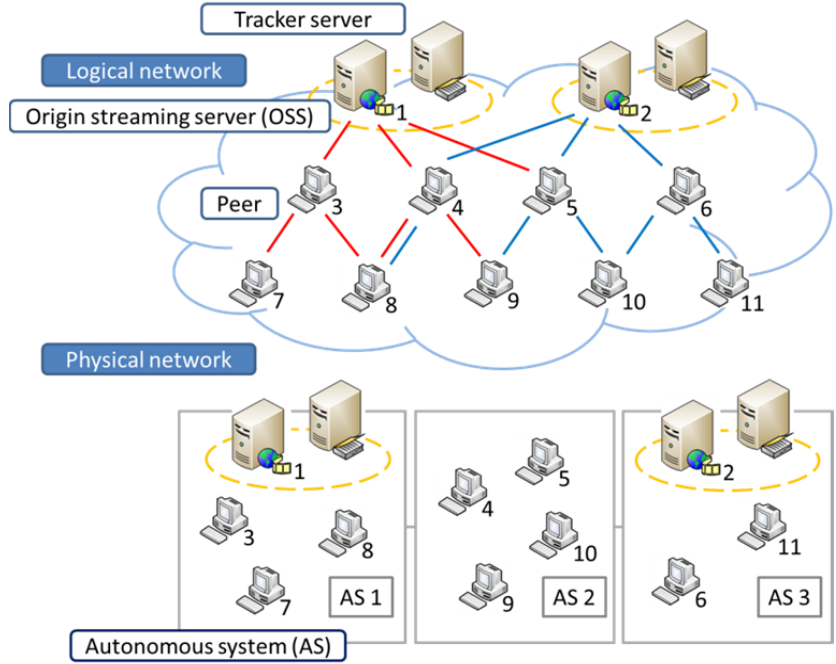

Figure 1. Streaming network model

A logical network represents a logical connection between nodes, and the number of hops between each node and OSS on a logical network is termed the logical hop count. The logical hop count of OSS is set to 0 and the maximum logical hop count is set to $H$. The logical hop count, $h_i^k$, of peer $i$ which requires video content $k$, is assumed to equal the maximum logical hop count in $\mathcal{P}^k(i)$ plus 1. If a peer selection algorithm cannot transmit sufficient data or the logical hop count exceeds $H$, the video quality will not be maintained and consequently a peer connection will not be established.

On the other hand, a physical network represents a physical connection between nodes, and the number of hops between nodes is termed the physical hop count. A physical hop count between node $i$ to node $j$, $d_{ij}$, is assumed to equal the hop count between ASs that contain node $i$ and $j$ plus 1.

For each node $i$, the transmission rate required for viewing video content $k$ is denoted as $N_i^k$, the effective bandwidth available for transmitting video to other nodes as $M_i$, and the transmission rate to node $j$ as $x_{i,j}^k$ (where $x_{i,j}^k = 0$, $j \in \mathcal{S}_0$). The whole rates $n_i^k$ and $m_i$ of receiving content and transmitting content by node $i$ satisfy

$$n_i^k = \sum_{j \in \mathcal{S}} x_{j,i}^k (N_i^k \geq n_i^k \geq 0), \tag{1}$$

and

$$m_i = \sum_{k \in \mathcal{K}} \sum_{j \in \mathcal{S}} x_{i,j}^k (M_i \geq m_i \geq 0), \tag{2}$$

respectively.

Video viewing time $W_i$ of each peer $i$ is assumed to follow an arbitrary distribution, and a peer that completes viewing video departs from the system. When peer $i$ departs from the system, each child peer $j \in \mathcal{C}^k(i)$ connected to a peer $i$ for video content $k$ will select a new transmission source from a reserved parent node set ($\mathcal{B}^k(i)$). For each peer $i$ for requiring video content $k$, a parent node set $\mathcal{P}^k(i)$ and a reserve node set $\mathcal{B}^k(i)$ are determined by the initial peer configuration algorithm. Note that a child node can has several parent nodes each of which provides with content. The total number of elements contained in $\mathcal{P}^k(i)$ and $\mathcal{B}^k(i)$ remains constant ($D = |\mathcal{P}^k(i) + \mathcal{B}^k(i)|$ for each $i$). If $\mathcal{P}^k(i)$ is altered by a reconnection, $\mathcal{B}^k(i)$ will be updated as necessary. However, if a peer cannot receive sufficient data to view video, it will depart from the system. After the departure, the waiting time until this peer demands service follows an exponential distribution with an average of $\lambda^{-1}$.

## 4. PROPOSED ALGORITHM

In this section, we propose an initial peer configuration algorithm to add service-demanding peers to a logical network. In the proposed initial peer configuration algorithm, a reserve node set for parents is concurrently determined to ensure that the P2P network has redundancy.

When using the algorithms explained in this section, understanding the topology between ASs is necessary. To this end, several methods have been proposed, including those using route information via traceroute and border gateway protocol (BGP) [15] and those using WHOIS database [16]. Mao et al. [17] have reported that a complete decision cannot be made using these methods, but some level of prediction is possible. Moreover, Fukushima et al. [8] have proposed peer configuration algorithms assuming the traffic between ASs cannot be perceived.

### 4.1. Peer Selection Algorithm

In the present study, we discuss a peer configuration algorithm based on MLH and MPH algorithms proposed in [7, 8]. Using these two algorithms allows the number of video viewers to be increased while reducing the physical traffic compared to a system where the network is constructed randomly.

#### 4.1.1. Minimum Logical Hop (MLH) Peer Selection Algorithm

If multiple peers, which each have a different logical hop count, are defined as transmission source peers, logical hop counts of child peers will be automatically determined in accordance with the parent peer that has the larger logical hop count, reducing the total number of connectable peers. Fukushima et al. [7, 8] have proposed the MLH algorithm, which can increase the number of peers capable of connecting to a network by repeating the procedure where the peer with the smallest logical hop count selects the transmission source peer. MLH algorithm explained below gives a parent node set $\mathcal{P}^k(i)$ for peer $i$ which requires video content $k$.

**MLH Peer Selection Algorithm:**

Step 1. Set $\mathcal{S}^A = \{j \in \mathcal{S}; \ M_j - m_j \geq N_i^k \ \text{and} \ h_j^k < h_i^k\}$.

Step 2. If it is satisfied that $\mathcal{S}^A = \emptyset$, then go to Step 5. Otherwise, define $\mathcal{S}^{A'}$ where each node, $j$, in $\mathcal{S}^{A'}$ is satisfied that

$$j = \operatorname*{argmin}_{j \in \mathcal{S}^A} h_j,$$

and set $\mathcal{S}^A = \mathcal{S}^A \setminus \mathcal{S}^{A'}$.

Step 3. Choose a node $l$ such that

$$l = \operatorname*{argmin}_{l \in \mathcal{S}^{A'}} d_{l,i}.$$

Set $\mathcal{P}^k(i) = \mathcal{P}^k(i) \cup \{l\}$ and $\mathcal{S}^{A'} = \mathcal{S}^{A'} \setminus \{l\}$.

Step 4. Set

$$x_{l,i}^k = \begin{cases} N_i^k - n_i^k, & \text{if } N_i^k - n_i^k \leq M_l - m_l, \\ M_k - m_k, & \text{otherwise,} \end{cases}$$

and update $n_i^k = n_i^k + x_{l,i}^k$, and $m_l = m_l + x_{l,i}^k$. If $n_i^k < N_i^k$, then go back to Step 3.

Step 5. If it is satisfied that $n_i^k = 0$, then the algorithm is successfully finished. Otherwise, reject the request for connection of peer $i$, that is, set $\mathcal{P}^k(i) = \emptyset$ and $x_{l,i}^k = 0$ for $l \in \mathcal{S}$.

### 4.1.2. Minimum Physical Hop (MPH) Peer Selection Algorithm

Construction of a logical network without considering a physical network may increase communication traffic volume. Because unnecessary inter-AS traffic may delay video data transmission and degrade video quality, the structures of the logical and physical networks should be as close as possible. Fukushima et al. [7, 8] have proposed the MPH algorithm, which aims to reduce the inter-AS traffic volume by connecting peers that have small physical hop counts to each other.

**MPH Peer Selection Algorithm:**

In the MLH algorithm, the logical hop count in Step 2 is replaced with the physical hop count from node $i$. Similarly, the physical hop count from node $i$ in Step 3 is replaced with the logical hop count.

### 4.2. Improvement of the Peer Selection Algorithms

### 4.2.1. Bandwidth-based Recursive Peer Exchange

To maximize the number of joining peers in a P2P network in which the logical hop count from OSS is restricted, it is important to locate peers with larger bandwidths in the higher-level layers of a logical network. Fukushima et al. [7, 8] did not consider this, and thus, the number of concurrently connected peers may decrease as time advances in a network where peers depart from the system. To overcome this issue, we propose a peer exchange algorithm by considering each peer's bandwidth. In this exchange algorithm, for each video content $k \in \mathcal{K}(i)$, $\mathcal{P}^k(i)$ and effective bandwidth $M_i$ are compared when peer $i$ becomes available to transmit data, and the locations of two peers are exchanged in a logical network if $M_i$ is larger than the bandwidth of a parent peer. By recursively applying this location exchange procedure, peers with larger bandwidths are consequently located in higher-level layers in the logical network.

Explained below seeks parent node set $\mathcal{P}^k(i)$ and child node set $\mathcal{C}^k(i)$ for peer $i$ that requests a connection. Herein a peer subject to exchange with peer $i$ is set to peer $j$.

**Peer Exchange Algorithm 1:**

Step1. Set

$$\mathcal{S}^{\mathrm{E}} = \{l \in \mathcal{P}^k(i); \ M_i > M_l \text{ and } m_i \leq m_l\},$$

and find $j = \underset{j \in \mathcal{S}^{\mathrm{E}}}{\mathrm{argmin}}\{M_j\}$. If it is satisfied that $\mathcal{S}^{\mathrm{E}} = \emptyset$, then finish the algorithm.

Step 2. Exchange the connections of node $i$ for those of node $j$ such that node $i$ is a parent node of node $j$. That is, the sets of parents and child nodes for node $i$ are $\mathcal{P}^k(j)$ and $\mathcal{C}^k(j)$, and those for $j$ are $\mathcal{P}^k(i)$ and $\mathcal{C}^k(i)$.

Step 3. If $\mathcal{P}^k(i) = \mathcal{S}_1 = \emptyset$, then finish the algorithm. Otherwise, go back to Step 1.

### 4.2.2. Physical Traffic-based Peer Exchange

The exchange algorithm in the previous section does not consider physical traffic; therefore, it may increase physical traffic when it is combined with topology construction algorithms such as the MPH algorithm. To overcome this issue, we propose an exchange algorithm to reduce physical traffic. To reduce the physical traffic, we first focus on the cost of a physical hop that arises upon the connection of a peer. Function $Z_i$, which gives physical hop counts, is defined by the weighted sum of the physical hop counts and the transmission volume for all connections of a peer $i$ as shown in the following equation.

$$Z_i = \sum_{k \in \mathcal{K}} \left\{ \sum_{j \in \mathcal{P}^k(i)} d_{j,i} x_{j,i}^k + \sum_{j \in \mathcal{C}^k(i)} d_{i,j} x_{i,j}^k \right\}, \tag{3}$$

The peer exchange algorithm explained below gives a parent node set $\mathcal{P}^k(i)$ and a child node set $\mathcal{C}^k(i)$ for peer $i$ which requiring video content $k$. Herein a peer subject to exchange with peer $i$ is set to peer $j$.

**Peer Exchange Algorithm 2:**

Step 1. Set

$$\mathcal{S}^{\mathrm{F}}(i) = \{j \in \mathcal{S}_1; \ M_j = M_i\}.$$

Step 2. For each node $j \in \mathcal{S}^{\mathrm{F}}(i)$, exchange it for node $i$ and compute the costs $Z'_i$ and $Z'_j$ from eq. (3).

Step 3. Find peer $j$ such that $Z_i + Z_j > Z'_i + Z'_j$ and that maximizes $(Z_i + Z_j) - (Z'_i + Z'_j)$.

Step 4. Exchange the connections of node $i$ for those of node $j$, that is, for each $k$ exchange the sets of parents and child nodes for node $i$ are $\mathcal{P}^k(j)$ and $\mathcal{C}^k(j)$, and those for $j$ are $\mathcal{P}^k(i)$ and $\mathcal{C}^k(i)$.

### 4.3. Initial Peer Configuration Algorithm

The initial peer configuration algorithm used in the present study is a combination of the peer selection algorithm proposed by Fukushima et al. [7, 8] and the peer exchange algorithm discussed in Section 4.2. The initial peer configuration algorithm, which calculates a parent node set $\mathcal{P}^k(i)$, a reserve node set $\mathcal{B}^k(i)$, and a child node set $\mathcal{C}^k(i)$ for a peer $i$, is explained below.

**Initial Peer Configuration Algorithm:**

Step 1. For each $k \in \mathcal{K}$, initialize the sets of parent, reserve and child nodes as $\mathcal{P}^k(i) = \emptyset$, $\mathcal{B}^k(i) = \emptyset$ and $\mathcal{C}^k(i) = \emptyset$, respectively.

Step 2. For each $k \in \mathcal{K}$, determine $\mathcal{P}^k(i)$ using **MLH** (or **MPH**) **peer selection algorithm**.

Step 3. For each $k \in \mathcal{K}$, set $\hat{\mathcal{P}}^k(i) = \mathcal{P}^k(i)$, and update $\mathcal{P}^k(i)$, $\mathcal{C}^k(i)$ using **peer exchange algorithm 1**.

Step 4. Update $\mathcal{P}^k(i)$, $\mathcal{C}^k(i)$ for each $k \in \mathcal{K}$ using **peer exchange algorithm 2**.

Step 5. For each $k \in \mathcal{K}$, choose a peer $j$ randomly (using a membership algorithm such as SCAMP), and $\mathcal{B}^k(i) = \mathcal{B}^k(i) \cup \{j\}$. Continue the above process as long as it is satisfied $|\mathcal{B}^k(i)| \leq D - |\mathcal{P}^k(i)|$.

## 5. SIMULATION EXPERIMENTS

In our simulation, we evaluate the number of joining peer and traffic load. For evaluating the traffic load of the peer-to-peer streaming service, we use the congestion degree [8].

The topology of the network (a Barabasi-Albert network [18] consisting of 15 ASs connected each other with 50 links) is generated with BRITE [19]. The network has one OSS and 30000 peers.

The effective bandwidth $M_i$ of each peer $i$, $i \in \mathcal{S}_1$ is uniformly distributed in the range from 0.5 [Mbps] to 10.0 [Mbps], and the effective bandwidth $M_i$ of the OSS $i$, $i \in \mathcal{S}_0$ is 30.0 [Mbps]. The transfer rates $N_i$, $i \in \mathcal{S}_1$ required for viewing video is 2.0 [Mbps], and the maximum number of logical hop count $H$ is 4. The number $K$ of video contents are 2, and the probabilities of peer requests are $P(\{\emptyset\}) = 0$ and $P(\{1\}) = P(\{2\}) = P(\{1,2\}) = 1/3$. The video viewing time $W_i$ is according to a log-normal distribution where the mean is 3 [hours] and the coefficients of variation is 6.

Denote by $n_i^k(t)$ and $x_{i,j}^k(t)$ the rate of receiving video content $k$ and the transmission rate of video content $k$ from node $i$ to $j$ at time $t$, respectively. We also denote by $T_d$ the seconds in a day (i.e., $T_d = 86400$). Then, the time averages of the number of joining peers and the congestion degree are given as follows:

$$\frac{1}{T_d} \int_0^{T_d} \sum_{k \in \mathcal{K}} \sum_{i \in \mathcal{S}_1} I_{\{n_i^k(t) = N_i^k\}} dt, \qquad (4)$$

and

$$\frac{1}{T_d} \int_0^{T_d} \frac{\sum_{k \in \mathcal{K}} \sum_{i \in \mathcal{S}} \sum_{j \in \mathcal{S}_1} x_{i,j}^k(t) d_{i,j}}{\sum_{k \in \mathcal{K}} \sum_{i \in \mathcal{S}} \sum_{j \in \mathcal{S}_1} x_{i,j}^k(t)} dt, \qquad (5)$$

where $I_A$ is the indicator function, that is, the value is 1 if $A$ is true, and otherwise the value is 0. In this simulation experiments, the simulation continues until the simulation time exceeds 12 days. We regard the time average of each performance measure in one day as one simulation batch. We implement the batch mean method (see e.g., [20]) for calculating the 95% confidence interval of each performance measure where the batch size is 10. Note that we discard the first two days since initial states are not always steady-states.

To show the effectiveness of our proposed algorithm, we compare with, SCAMP [6] and peer selection algorithms proposed by Fukushima et al. [8].

Figure 2 shows the number of joining peers of our simulation experiments for various values of mean waiting time $\lambda^{-1}$. From Figure 2, our algorithms give the better performance than SCAMP, and the MLH, MPH algorithms proposed by Fukushima et al [8]. Especially, it is effective in the performance of P2P streaming services when it is combined with the minimum logical hop peer selection algorithm. This improvement may be due to the benefit of the bandwidth-based recursive peer exchange algorithm.

Figure 3 shows the congestion degree of our simulation experiments for various values of mean waiting time $\lambda^{-1}$. From Figure 3, we observe that our algorithms reduce the network load by around $10 - 30\%$ as compared to other algorithms.

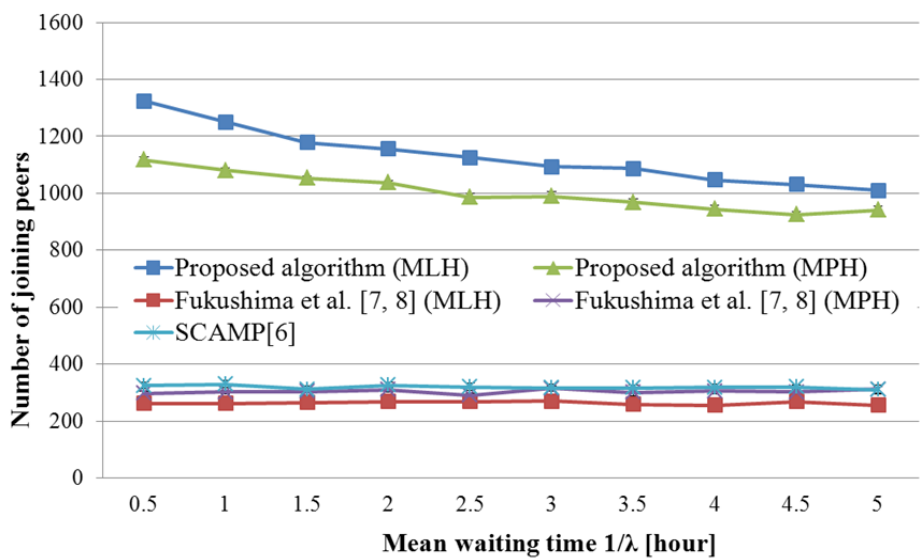

Figure 2. Number of joining peers for each value of mean waiting time

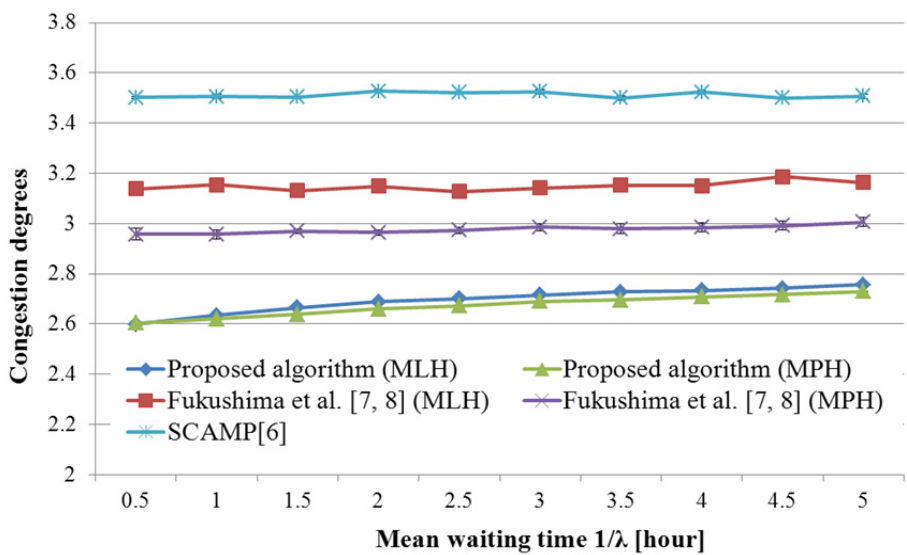

Figure 3. Congestion degrees for each value of mean waiting time

# 6. CONCLUSION

In this paper, we have proposed an initial peer configuration algorithm in a multi-streaming peer-to-peer (P2P) streaming network. The proposed algorithm has been based on a mesh-pull approach where any node has multiple parent and child nodes as neighboring nodes, and content transmitted between these neighboring nodes depends on their parent-child relationships. The main algorithm consists of three sub-algorithms each of which independently improves the performance of the network. Through some simulation experiments, we have shown that our algorithm outperforms the existing peer configuration algorithms.

Some problems remain to be solved. For example, we must improve our algorithm to avoid failure of re-connection for child nodes due to a parent peer leaving. We can also consider the effect of the cross traffics on a physical network on the system performance.

**Authors**

Short Biography

Tomoyuki Ishii received the Bachelor of Engineering from Kanagawa Institute of Technology.

He is currently a student of Master at Kanagawa Institute of Technology.

His research interests include, computer and communication network, peer-to-peer network, and performance evaluation.

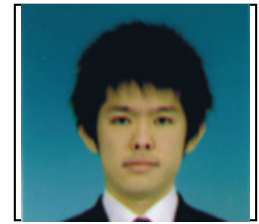

Atsushi Inoie received the Ph.D degree in Systems and Information Engineering from University of Tsukuba, Tsukuba Science City, Ibaraki, Japan, in 2006.

He is currently an Assistant Professor at Kanagawa Institute of Technology, Atsugi-city, Kanagawa, Japan.

His research interests include queueing theory, optimization theory, game theory, multimedia communications, wireless communication and distributed systems.

He is a member of the IEEE.

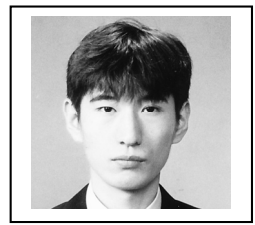